\documentclass[preprint,showpacs,showkeywords,preprintnumbers,amsmath,amssymb]{revtex4}

\usepackage{graphicx}
\usepackage{dcolumn}
\usepackage{bm}
\begin{document}

\title{Entropy and Its Quantum Thermodynamical Implication for Anomalous
Spectral Systems}

\author{Chun-Yang Wang$^1$\footnote{Corresponding author. Electronic mail:
wchy@mail.bnu.edu.cn}\footnote{Researcher in the Physical
Post-doctoral Circulation Station of Qufu Normal University}}
\author {An-Qi Zhao$^1$}
\author {Xiang-Mu Kong$^1$}
\author {Jing-Dong Bao$^2$$^,$$^3$}
\affiliation{$^1$Shandong Provincial Key Laboratory of Laser
Polarization and Information Technology, Department of Physics, Qufu
Normal University, Qufu 273165, China\\$^2$Department of Physics,
Beijing Normal University, Beijing 100875, China\\$^3$Center of
Theoretical Nuclear Physics, National Laboratory of Heavy-ion
Accelerator, Lanzhou 730000, China}


\begin{abstract}
The state function entropy and its quantum thermodynamical
implication for two typical dissipative systems with anomalous
spectral densities are studied by investigating on their
low-temperature quantum behavior. In all cases it is found that the
entropy decays quickly and vanishes as the temperature approaches
zero. This reveals a good conformity with the third law of
thermodynamics and provides another evidence for the validity of
fundamental thermodynamical laws in the quantum dissipative region.
\end{abstract}


 \maketitle

\section{INTRODUCTION}

The widespread interest on dissipative environments in recent years
has highlighted the critical role that it plays in the study of
mesoscopic systems \cite{sec10}, statistical mechanics
\cite{sec11,sec12}, quantum computation and fundamental quantum
physics \cite{sec13,sec14}. In particular, the quantum thermodynamic
properties of such stochastic coupling open systems has long been a
hot spot of chemistry and physics \cite{sec15,sec16}. Whereas many
new facets of old results have emerged bringing out lots of subtle
issues. It is important to be cautious to question the validity of
fundamental laws especially the three laws of thermodynamics.

Of all the fundamental laws in thermodynamics, the third law which
is contributed by W. H. Nernst in the critical analysis of various
chemical and electrochemical reactions bears prominent consequences
for quantum mechanics and low-temperature physics studies
\cite{sec17,sec18,sec19}. It confirmed the fact that the absolute
zero temperature is unattainable because of the immediate
coincidence between the isotherm and the isentrope (adiabat)
\cite{sec19}. Therefore the Carnot engine can never reach $100\%$
efficiency for any finite temperature.

Great progress in thermodynamics attributed to this law has been
witnessed in curing the known deviations by quantum statistical
mechanics and interactions among particles according to common
wisdom, although some unfulfillment may still exist. Intriguingly it
has been found that finite dissipation helps to ensure the third law
rather than deviate it \cite{PHan1,PHan2}. Further investigations
sequentially illuminated the predominance of anomalous couplings in
assuring the third law \cite{chyw}. However, the mechanism of how
quantum dissipation influences the low-temperature thermodynamic
properties of the system is still an important subject needs to be
clarified. Meanwhile, the recent widespread interest in the
low-temperature behavior of small systems has provided a new point
of viewing the pivotal role that dissipative environment plays in a
virginal physical field of study, namely, quantum thermodynamics,
for which the validity of the third law is an unavoidable subject to
be elucidated.

Basing on these considerations, in this paper we devote our mind to
the state function entropy of two typical quantum dissipative
systems with anomalous spectral density, dedicating to elucidate its
properties of evolving with temperature and quantum thermodynamical
implication. In Sec. \ref{sec2}, the remarkable integral method
which is to be used in getting the free energy and entropy is
briefly retrospected. Sec. \ref{sec3} gives the explicit expression
of entropy for two typical anomalous spectral systems. Sec.
\ref{sec4} serves as a summary of our conclusion in which the
implicit application of this work is deliberated as well.

\section{integral method}\label{sec2}

In an earlier paper by G.W. Ford, et al. \cite{sec20} a remarkable
integral method was presented which provided us a convenient
shortcut to calculate the free energy and entropy of a complex
dissipative system \cite{sec21,fre1,fre2,fre21}. The primary
procedure is to begin with writing the free energy of the quantum
oscillator as
\begin{eqnarray}
F(T)=\frac{1}{\pi}\int^{\infty}_{0}d\omega
f(\omega,T)\textrm{Im}\left\{
\frac{\textrm{d}\log\alpha(\omega+i0^{+})}{\textrm{d}\omega}\right\}\label{eq:fre},
\end{eqnarray}
where $f(\omega,T)$ is the free energy of a single oscillator of
frequency $\omega$, given by
$f(\omega,T)=k_{B}T\log[1-\exp(-\hbar\omega/k_{B}T)]$ with the
zero-point contribution $\hbar\omega/2$ being omitted. While
$\alpha(\omega)$ denotes the generalized susceptibility which can be
got from the corresponding equation of motion of the system. After
accomplishing this integration, the entropy of the quantum
oscillator can then easily be got from a single derivation as
\begin{eqnarray}
S(T)=-\frac{\partial F(T)}{\partial T}.\label{eq.entropy}
\end{eqnarray}

Since the function $f(\omega,T)$ in Eq.(\ref{eq:fre}) vanishes
exponentially for $\omega\gg k_{B}T/\hbar$, the total integrand is
then confined only to low frequencies as $T\rightarrow0$ and the
free energy together with the entropy then can be calculated by
expanding the factor multiplying $f(\omega,T)$ in the powers of
$\omega$. So what is essentially left to be determined in the
following is the generalized susceptibility $\alpha(\omega)$ which
can be resulted from analytically solving the quantum Langevin
equation (QLE)
\begin{eqnarray}
M\ddot{x}+M\int^{t}_{0}dt'\gamma(t-t')\dot{x}(t')+\partial_{x}U(x)=\xi(t),\label{eq.GLE}
\end{eqnarray}
where $\gamma(t)$ is the memory friction function and $\hat{\xi}(t)$
is the random force operator with zero mean, its correlation obeys
the quantum fluctuation-dissipation theorem \cite{sec15,nonOhmic}
\begin{eqnarray}
\langle \xi(t)\xi(t')\rangle_{s}=\frac{\beta\hbar}{\pi
}\int^{\infty}_{0} d\omega
J(\omega)\textrm{coth}(\frac{\beta\hbar\omega}{2})\textrm{cos}(t-t'),
\end{eqnarray}
where $\langle \cdots\rangle_{s}$ denotes the quantum symmetric
average operation and $\beta=1/k_{B}T$ is the inverse temperature.

In the case of harmonic potential
$U(x)=\frac{1}{2}M\omega^2_0x^{2}$, the QLE is linear and its
solution can be obtained by Fourier transformation as
\begin{eqnarray}
\tilde{x}(\omega)=\alpha(\omega)\tilde{\xi}(\omega),\label{result}
\end{eqnarray}
where $\tilde{x}(\omega)=\int_{-\infty}^{\infty}dtx(t)\exp(i\omega
t)$ and similarly noting is true for $\tilde{\xi}(\omega)$. This
results in the generalized susceptibility
\begin{eqnarray}
\alpha(\omega)=\left[-M\omega^{2}-iM\omega\tilde{\gamma}(\omega)+M\omega_{0}^{2}\right]^{-1}\label{eq:alpha}.
\end{eqnarray}

\section{entropy and Its Quantum Thermodynamical Implication} \label{sec3}

Essentially, for arbitrary systems the expression of the generalized
susceptibility $\alpha(\omega)$ can be easily obtained given the
Fourier transform of the memory friction function is known. For
example, supposing $\tilde{\gamma}(\omega)$ can be written in the
following form of complex
\begin{eqnarray}
\tilde{\gamma}(\omega)=\textrm{Re}[\tilde{\gamma}(\omega)]+i\hspace{0.05cm}\textrm{Im}[\tilde{\gamma}(\omega)],\label{memo-fric}
\end{eqnarray}
then after some algebra, we will obtain in the low-frequency limit
\begin{eqnarray}
&&\textrm{Im}\left\{\frac{\textrm{d}\log\alpha(\omega)}{\textrm{d}\omega}\right\}\nonumber\\
&&\hspace{0.5cm}=\frac{\left[\omega_{0}^{2}\left(1+\omega
\frac{\textrm{d}}{\textrm{d}\omega}\right)+\omega^{2}\left(1+\textrm{Im}[\tilde{\gamma}(\omega)]-\omega\frac{\textrm{d}}{\textrm{d}\omega}\right)\right]\textrm{Re}[\tilde{\gamma}(\omega)]}
{(\omega_{_{0}}^{2}-\omega^{2})^{2}+\omega^{2}\tilde{\gamma}(\omega)^{2}}\nonumber\\
&&\hspace{0.5cm}\cong\frac{(1+\omega
\frac{\textrm{d}}{\textrm{d}\omega})\textrm{Re}[\tilde{\gamma}(\omega)]}{\omega_{0}^{2}}.\label{eq:imgs}
\end{eqnarray}
Since it has been confirmed in previous studies for the quantum
dissipative systems \cite{sec15,fre1,cl1}
\begin{eqnarray}
\textrm{Re}[\tilde{\gamma}(\omega)]=\frac{1}{M}\frac{J(\omega)}{\omega},\label{eq:Re}
\end{eqnarray}
this provides a fundamentally convenient way to investigate the
thermodynamical properties of the system via the spectral density
$J(\omega)$ in case of not knowing the explicit form of
$\tilde{\gamma}(\omega)$. Here in the following, the state function
entropy of two typical quantum dissipative systems with anomalous
spectral is studied using above-mentioned relations.

\subsection{Drude system}

\begin{figure}
\includegraphics[scale=0.8]{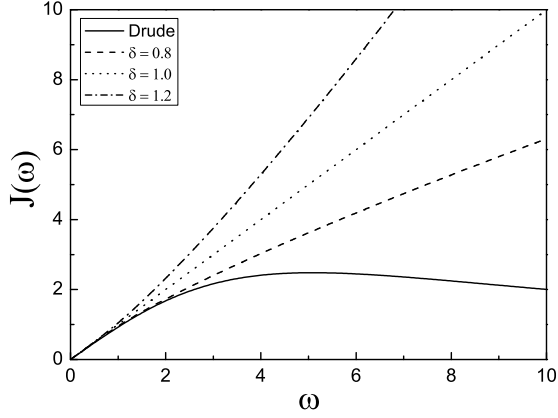}
\caption{The spectral density of the Drude and non-Ohmic (e.g.
$\delta=0.8$ and $\delta=1.2$) system as functions of the frequency
compared with the Ohmic case ($\delta=1.0$), where dimensionless
parameters such as $M=1.0$ and
$\omega_{r}=\gamma_{_{\delta}}=\gamma=1.0$ are used while
$\omega_{_{d}}=5.0$.\label{spectral}}
\end{figure}

Firstly, let us consider the particular case of the system with a
frequency-dependent Drude type of spectral density
\cite{sec15,DrudeS}
\begin{eqnarray}
J(\omega)=M\gamma\omega\frac{1}{1+\omega^{2}/\omega_{_{d}}^{2}},\label{eq:DS}
\end{eqnarray}
where $\gamma$ is the friction constant and $\omega_{_{d}}$ a Drude
cut-off frequency. Seen in Fig.\ref{spectral}, when the relevant
frequencies of the system are much lower than $\omega_{_{d}}$, the
reservoir described by Eq.(\ref{eq:DS}) behaves like an Ohmic bath
with constant effective damping strength $\gamma$. The Drude model
is therefore always be treated as a type of generalized Ohmic
spectral.

After some algebra we obtain in the low-frequency limit
\begin{eqnarray}
&&\textrm{Im}\left\{\frac{\textrm{d}\log\alpha(\omega)}{\textrm{d}\omega}\right\}=
\frac{\gamma}{\omega_{0}^{2}}\left[\frac{1-\omega^{2}/\omega_{_{d}}^{2}}{(1+\omega^{2}/\omega_{_{d}}^{2})^{2}}\right]\cong\frac{\gamma}{\omega_{0}^{2}}.\label{eq:Drudeim}
\end{eqnarray}
Hence, we get the expression of the free energy of quantum
oscillator at low temperature,
\begin{eqnarray}
F(T)&=& \frac{\gamma k_{B}T}{\pi\omega_{_{0}}^{2}}
\int^{\infty}_{0}d\omega\log[1-\exp(-\hbar\omega/k_{B}T)]\nonumber\\&
=&-\frac{\pi}{6}\hbar\gamma\left(\frac{k_{B}T}{\hbar\omega_{_{0}}}\right)^{2}.
\end{eqnarray}
The entropy thus reads
\begin{eqnarray}
S(T)=-\frac{\partial F(T)}{\partial
T}=\frac{\pi}{3}\gamma\left(\frac{k_{B}^{2}T}{\hbar\omega_{_{0}}^{2}}\right).
\end{eqnarray}
Seen from it, the large-frequency cut-off of the Drude spectral is
revealed to have no influence on the low-temperature properties of
the quantum dissipative system. A kind of linear decaying entropy
identical to the case of typical Ohmic friction is obtained. As
$T\rightarrow0$, $S(T)$ vanishes principle to $T$, in perfect
conformity with the third law of thermodynamics.

\subsection{non-Ohmic system}

In order to have more intensive understanding on the system's
low-temperature quantum properties, we turn in the following to the
non-Ohmic damping system which is another type of dissipative
systems with anomalous spectral.

The non-Ohmic type of spectral emerges in various frameworks in
physics \cite{sec15,nonO2,nonO3}. For example, quantized chaotic
systems might exhibit non-Ohmic fluctuations due to semi-classically
implied long time power-law correlations \cite{nonO4,nonO5,nonO6}.
Other examples may appear in the context of a many-particle system
\cite{nonO7,nonO8}, where the hierarchy of states and associated
couplings, ranging from the single-particle levels to the
exponentially dense spectrum of complicated many particle
excitations, can lead to a very structural non-Ohmic band profile
describing the residual interactions. As is seen in Fig.
\ref{spectral}, in any cases the typical spectral is not like
``white noise''.

In general, the non-Ohmic spectral density is written in the
following form
\begin{eqnarray}
J(\omega)=M\gamma_{_{\delta}}\left(\frac{\omega}{\omega_{r}}\right)^{\delta},\label{eq:NOS}
\end{eqnarray}
where $\delta$ is the power exponent taking values between 0 and 2,
$\gamma_{_{\delta}}$ is the symmetrical friction constant tensor,
and $\omega_{r}$ denotes a reference frequency allowing for
$\gamma_{_{\delta}}$ to have the dimension of a viscosity at any
$\delta$.

\begin{figure}
\includegraphics[scale=0.8]{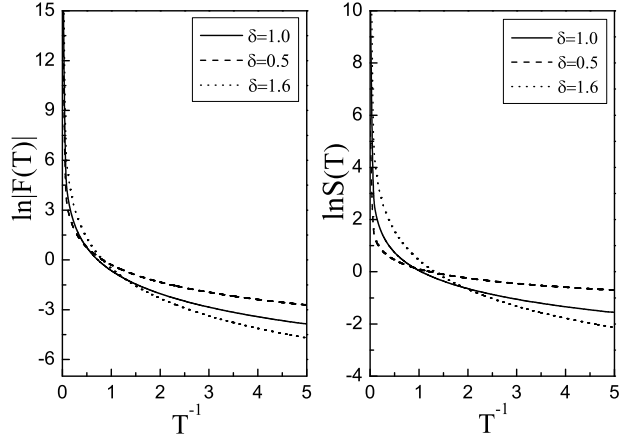}
\caption{The free energy and entropy of the Drude ($\delta=1.0$) and
non-Ohmic (e.g. $\delta=0.5$ and $\delta=1.6$) system as functions
of the inverse temperature. Where dimensionless parameters such as
$\hbar\omega_{_{r}}=k_{B}=\gamma_{_{\delta}}=1.0$ as well as
$\omega_{_{0}}=1.0$ are used.\label{non-OFE}}
\end{figure}

In the low-frequency limit we have
\begin{eqnarray}
\textrm{Im}\left\{\frac{\textrm{d}\log\alpha(\omega)}{\textrm{d}\omega}\right\}=
\frac{\delta\gamma_{_{\delta}}\omega^{\delta-1}}{\omega_{0}^{2}\omega_{r}^{\delta}}.\label{eq:nonOhmicim}
\end{eqnarray}
Hence, we get the expression of the free energy at low temperature
\begin{eqnarray}
F(T)&=&\frac{\delta\gamma_{_{\delta}}k_{B}T}{\pi\omega_{0}^{2}\omega_{r}^{\delta}}
\int^{\infty}_{0}d\omega\omega^{\delta-1}\log[1-\exp(-\hbar\omega/k_{B}T)]\nonumber\\
&=&-\Gamma(\delta+1)\zeta(\delta+1)\frac{\gamma_{_{\delta}}\hbar\omega_{r}}
{\pi\omega_{0}^{2}}\left(\frac{k_{B}T}{\hbar\omega_{r}}\right)^{\delta+1}.\label{eq:nonOhmicfe}
\end{eqnarray}
The entropy thus reads
\begin{eqnarray}
S(T)=\Gamma(\delta+2)\zeta(\delta+1)\frac{\gamma_{_{\delta}}k_{B}}
{\pi\omega_{0}^{2}}\left(\frac{k_{B}T}{\hbar\omega_{r}}\right)^{\delta},
\end{eqnarray}
where the gamma function
$\Gamma(z)=\int^{\infty}_{0}t^{z-1}e^{-t}dt$ and the Riemann's
zeta-function $\zeta(z)=\Sigma_{n=1}^{\infty}\frac{1}{n^{z}}$ is
used. With this result we conclude again that $S(T)\rightarrow0$ as
$T\rightarrow0$ in agreement with the third law as is shown in
Fig.\ref{non-OFE}. In all cases the entropies are revealed to decay
as a $\delta$ order power function of the temperature. This is much
different from that of the Drude system as well as the general Ohmic
case. Formally it results from the fact that
Eq.(\ref{eq:nonOhmicim}) has a factor $\omega^{\delta-1}$ whereas
the corresponding result of Eq.(\ref{eq:Drudeim}) is independent of
the frequency. But actually its essential reason lives in the fact
that the coupling strength between the system and the reservoir of
these two damping systems is completely different from each other.

\begin{figure}
\includegraphics[scale=0.7]{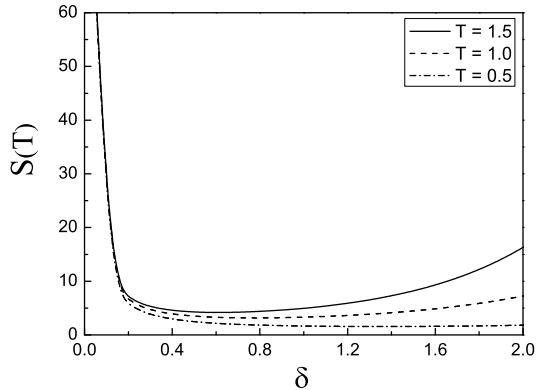}
\caption{The entropy of the non-Ohmic system as functions of the
friction exponent $\delta$ at various $T$. Where dimensionless
parameters such as $\hbar\omega_{_{r}}=k_{B}=\gamma_{_{\delta}}=1.0$
as well as $\omega_{_{0}}=1.0$ are used.\label{non-Oentropy}}
\end{figure}

For more detailed information, we plot in Fig.\ref{non-Oentropy} the
entropy of the non-Ohmic systems as functions of the friction
exponent $\delta$ at different $T$. From which we can see that the
entropy of the non-Ohmic systems evolves non-monotonically only if
the system temperature is very low. This reveals a common influence
of the non-Ohmic damping on the quantum thermodynamic properties of
the system. Comparing $S(T)$ in the sub-Ohmic $(0<\delta<1)$ and
super-Ohmic $(1<\delta<2)$ range with that of the Ohmic case
$(\delta=1)$ one can found not all the non-Ohmic damping case is
beneficial to the quantum dissipative system. Because the
low-temperature quantum behavior tends to be annihilated in the
limit of sub- or super-Ohmic range of damping. This is a particular
conclusion resulted only from the study of non-Ohmic damping
systems.

\section{summary and discussion} \label{sec4}

In summary, we have studied in this paper the low-temperature
thermodynamical properties of two typical anomalous quantum
dissipative systems by deriving the entropy function from the
spectral density. In all cases it is found that the entropy decays
quickly and vanishes as the temperature approaches zero in good
conformity with the third law of thermodynamics. The results
obtained in this study provides another evidence for the validity of
fundamental laws in the quantum dissipative region and may turn out
to be relevant to experiments in nanoscience where one always tests
the quantum thermodynamics of small systems coupled to a heat bath.
Experimentally, this work may also provide useful information for
some other studies such as those in connection with the radiation of
black-body.

\section * {ACKNOWLEDGEMENTS}

This work was supported by the Shandong Province Science Foundation
for Youths (Grant No.ZR2011AQ016), the Shandong Province
Postdoctoral Innovation Program Foundation (Grant No.201002015), the
Scientific Research Starting Foundation and Youth Foundation of Qufu
Normal University (Grant No.XJ201009).

\end{document}